\documentclass[prl,preprint,superscriptaddress]{revtex4}
\usepackage{graphicx}
\usepackage{dcolumn}
\usepackage{amssymb}
\usepackage{bm}
\usepackage{pifont}
\usepackage{epsfig}
\usepackage{psfrag}
\usepackage{epstopdf}
\usepackage[usenames]{color}
\begin{document}

\title{Shape of Dynamical Heterogeneities and the Stokes-Einstein and Stokes-Einstein-Debye Relations in Suspensions of Colloidal Ellipsoids}
\author{Chandan K Mishra}\thanks{To whom correspondence should      
be addressed} \email{mishrachandan23@gmail.com}
\affiliation{Chemistry and Physics of Materials Unit, Jawaharlal Nehru Centre for Advanced Scientific Research, Jakkur, Bangalore - 560064, INDIA}
\author{Rajesh Ganapathy}
\affiliation{International Centre for Materials Science, Jawaharlal Nehru Centre for Advanced Scientific Research, Jakkur, Bangalore - 560064, INDIA}
\date{\today}
\draft
\begin{abstract}
We examine the role of shape of dynamical heterogeneities on the validity of the Stokes-Einstein (SE) and Stokes-Einstein-Debye (SED) relations in quasi-two-dimensional suspensions of colloidal ellipsoids. For ellipsoids with repulsive interactions, although the orientational relaxation time remains coupled to the structural one, the SED relation by the Einstein formalism shows a breakdown. Strikingly, we find that it is the change in the shape of the dynamical heterogeneities from string-like to compact and not just their presence that results in the breakdown of both the SE and SED relations. On introducing a short-range depletion attraction between the ellipsoids, associated with the lack of morphological evolution of dynamical heterogeneities, the SE and SED relations remain valid even for deep supercooling. Our observations are consistent with numerical predictions.

\end{abstract}
\maketitle
\renewcommand{\thefootnote}

While it is not possible to distinguish conventional liquids from supercooled ones using static structural measures, a dynamical signature unique to the latter is the presence of spatial and temporal heterogeneities \cite{DH_1, DH_2, DH_3, DH_4, DH_5}. Regions of predominantly fast particles that contribute primarily to diffusivity, $D$,  are spatially decoupled from the slow regions that govern the bulk viscosity $\eta$ or equivalently the structural relaxation time $\tau_\alpha$ \cite{DH_4, DH_5}. A consequence of dynamical heterogeneities (DH) is the Stokes-Einstein (SE) relation, $D^{T} = {k_{B}T\over{6\pi\eta a}}$, \cite{einstein} and/or the Stokes-Einstein-Debye (SED) relation, $D^{\theta} = {k_{B}T\over{8\pi\eta a^3}}$, \cite{debye} that are hallmarks of simple liquids, breakdown \cite{DH_4, DH_5, breakdown_papers}. Here, the superscripts $T, \theta$ denote translational and rotational degrees of freedom (DOF), respectively, $a$ is the particle radius and $k_{B}T$ is the thermal energy. Numerous studies have explored the connections between the extent of SE/SED breakdown and the standard quantifiers of DH, namely, the stretching exponent $\beta$, the non-Gaussian parameter, $\alpha_2(t)$ and the dynamic susceptibility $\chi_4$ \cite{DH_4, breakdown_quantifiers_1, breakdown_quantifiers_3, breakdown_quantifiers_2}. It has also been shown recently that the extent of SE breakdown depends on the dimensionality of the space inhabited by the liquid - with the breakdown occurring at all temperatures, and is the strongest, in two-dimensions (2D) and becoming progressively weaker with increasing dimensionality \cite{breakdown_quantifiers_2, dimension_dependence_1}. A key question, however, has remained unanswered. \textit{Do morphological changes in DH influence the breakdown of SE and/or SED relations?} \cite{szamel}. This question becomes all the more relevant in the context of the random first-order transition (RFOT) theory, a prominent thermodynamic theory of the glass transition, which predicts a change in the morphology of DH from string-like to compact on approaching the glass transition \cite{RFOT_cluster_shape_1, RFOT_cluster_shape_2}. In fact, recent simulations have suggested that although DH emerges at the onset temperature of slow dynamics $T_o$, the violation of the SE relation starts at the dynamical crossover temperature $T_s < T_o$, where the shape of DH also undergoes marked changes \cite{szamel}. Furthermore, recent colloid experiments have shown that DH go from string-like to compact on turning on attractive interactions \cite{yodh_prl}, albeit its influence on the breakdown of SE relation has not been explored. In fact, at present it is \textit{even} unclear whether DH  in the rotational DOF also shows morphological changes on approaching the glass transition, let alone its influence on the validity of the SED relation.

Before addressing the above questions, it is imperative to first highlight the lack of consensus between the two complementary approaches used to investigate the validity of SED relation in supercooled liquids. Numerical studies, where $D^\theta$ is directly extracted from particle trajectories - the `Einstein Method', find that the SED relation breaks down to the same extent as SE \cite{stanley}. On the contrary, studies which have access to the $n^{\text{th}}$-order orientational relaxation time $\tau_n$, only $n = 2$ can be accessed in molecular experiments, invoke the `Debye Model', $D^\theta \propto {1 \over {\tau_n}}$, and find the SED relation to be valid even for deep supercooling \cite{DH_2, SED_valid_Debye}. In supercooled liquids, the orientational correlators decay as stretched-exponentials leading to the failure of the Debye Model. Simulations on hard dumbbells find that $1 \over \tau_2$, nevertheless, continues to scale linearly with $\tau_\alpha$, albeit $D^\theta$ extracted from the Einstein method shows a breakdown \cite{eta_tau_alpha_1}. Recent colloid experiments that probed the dynamics of anisotropic tracers in a bath of smaller hard spheres, however find that the SED relation remains valid even close to glass transition irrespective of the method used \cite{weeks}. It is well-known that the breakdown of the SED/SE relation depends on the size, shape and roughness of the tracers with respect to the host \cite{breakdown_quantifiers_1, tracers}. It would be therefore be worthwhile to investigate the SED and SE relation in an experimental model system where particle self-diffusivities, as opposed to tracer diffusivities, can be directly accessed.

Suspensions of micrometer-sized colloidal ellipsoids are an ideal test bed to probe translational and rotational dynamics in real-space and with single-particle resolution. In this Letter, we use previously acquired microscopy data \cite{our_prl} to investigate the breakdown of the SE and SED relations in quasi-two-dimensional suspensions of colloidal ellipsoids, aspect ratio $\alpha = 2.1$, with repulsive as well as attractive interactions.  We show that consistent with RFOT theory, DH for both orientational and translational degrees of freedom become increasingly compact on approaching the glass transition. We find that the Debye Model fails irrespective of the nature of the inter-particle interactions. However, $1\over\tau_n$ couples with $\tau_\alpha$ for the repulsive case and decouples with it for the attractive case due to the onset of pseudonematic domains. Most importantly, we forge a direct link between the morphological evolution of DH and the breakdown of both the SED and SE relations.   

\begin{figure}[htbp]
\includegraphics[width=0.8\textwidth]{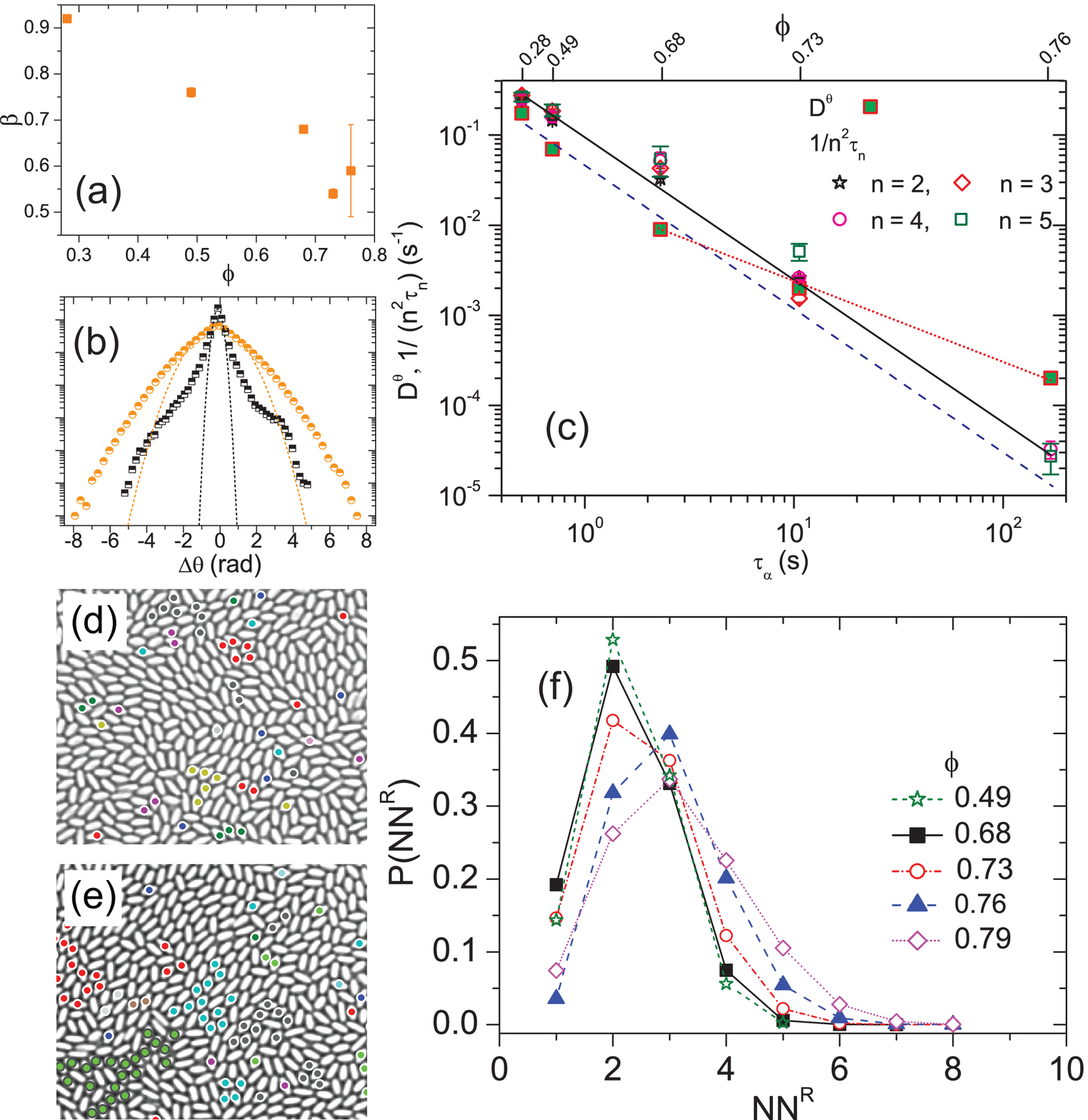}
\caption{(Color online). (a) Variation of the stretching exponent, $\beta$, obtained from fits to $L_2(t)$, with $\phi$. (b) Distribution of $\Delta \theta$, over $t^*$ for $\phi = 0.49$ (hall-filled circles) and $\phi = 0.76$ (half-filled squares). The dotted lines in (b) are Gaussian fits to $P(\Delta \theta)$. (c) Orientational diffusion coefficient, $D^{\theta}$, and inverse of the $n^{th}$-order orientational relaxation time, $1/ {n^2 \tau_n}$, versus the structural relaxation time $\tau_{\alpha}$. The lines in (c) show $\tau_\alpha^{-1.4}$ (solid and dashed lines) and $\tau_\alpha^{-0.9}$  (dotted line) dependencies. Clusters of top $10\%$ orientationally most-mobile particles for (d) $\phi = 0.76$, (e) $\phi = 0.79$. In (d) and (e), the colors correspond to distinct clusters. (f) Distribution of orientationally fast nearest-neighbours for a orientationally fast particle $P(NN^{\theta})$ for different $\phi$s.}
\label{Figure1}
\end{figure}

The experimental details are as described in \cite{our_prl, supplement} and we will not dwell on it here. First, we examined the validity of the SED relation, using the two approaches described earlier, for ellipsoids with purely repulsive interactions. To ascertain if the `Debye Model' can be used to estimate $D^\theta$, we computed the $n^{\text{th}}$-order orientational correlation function, $L_n(t) \equiv {{1 \over N}\langle\sum_{k=1}^{N}\text{cos}n(\Delta {\theta}_k(t))\rangle}$ \cite{our_prl, yhan}, for $n = 2..5$, and for all area fractions, $\phi$, investigated. Here, $\Delta \theta$ is the angular displacement of the $k^{th}$ ellipsoid, $t$ is the lag time and $\langle ... \rangle$ represents the time averaging. As seen in earlier experiments on colloidal ellpsoids ($\alpha = 6,9$) \cite{yhan}, even for $\phi = 0.28$, the long-time decay of $L_n(t) = \exp [-(t/{\tau_n})^{\beta}]$ was found to be a stretched-exponential ($\beta < 1$)  (Fig. \ref{Figure1}a). On approaching the glass transition area fraction $\phi^\theta_g = 0.80$, and in concord with findings from experiments and simulations \cite{breakdown_quantifiers_1, breakdown_quantifiers_3, eta_tau_alpha_1,our_prl, yhan, bagchi}, $\beta$ was found to decrease (Fig. \ref{Figure1}a). This clearly signals a growing departure from the simple Debye type relaxation dynamics and is consistent with earlier observations of the increase in the size of DH in the rotational DOF on approaching $\phi_g$ \cite{our_prl}. The presence of DH is also reflected in the non-Gaussian nature of particle displacements evaluated over the cage rearrangement time $t^*$ (Fig. \ref{Figure1}b). 

Next, following the Einstein Method, we directly evaluated $D^\theta$ from the long-time diffusive region of the mean-squared angular displacements, $ \langle \Delta \theta^2(t) \rangle = 2D^{\theta} t$ (see the Supplemental Material \cite{supplement}). We were unable to calculate $D^\theta$ for $\phi > 0.76$, since $ \langle \Delta \theta^2(t) \rangle$ does not reach the diffusive limit. Figure \ref{Figure1}c shows $1 \over{\tau_n}$, for $n = 2..5$, and $D^\theta$ plotted against $\tau_\alpha$. $\tau_\alpha$ at various $\phi$'s was obtained from the decay of the self-intermediate scattering function, ${F_s (q, t)} \equiv {{1 \over N}\langle\sum_{k=1}^{N}\text{exp} [i{\bf q}\cdot \Delta \textbf{r}_k(t)]\rangle}$ to $1\over e$. Here the wavevector $q$ is chosen to correspond to the first peak of the pair-correlation function since for this particular choice, $\tau_\alpha$ is known to precisely map changes in $\eta$ \cite{eta_tau_alpha_2}. Owing to the failure of the Debye Model even at low $\phi$'s, $1\over {n^2\tau_n}$ and $D^\theta$ do not collapse although they scale similarly with $\tau_\alpha$ for $\phi \leq 0.68$. At low $\phi$'s, $D^\theta$ and $1\over{n^2\tau_n}$ are found to scale as ${\tau_\alpha}^{-\xi}$, with $\xi > 1$ (Fig. \ref{Figure1}c). Although, physical insights into these observations is lacking, our findings are consistent with simulation results \cite{breakdown_quantifiers_2, horrowell}.  Most strikingly, while $D^\theta$ shows clear signatures of a decoupling with $\tau_\alpha$ for $\phi > 0.68$, $1 \over {n^2\tau_n}$ for all $n$ show complete collapse and stay coupled to $\tau_\alpha$ (Fig. \ref{Figure1}c). These results are in agreement with recent simulations on hard dumbbells \cite{eta_tau_alpha_1}. Further, analogous to $\tau_\alpha$, $\tau_n$ is also dominated by the dynamics of slow particles. Since for $\alpha = 2.1$ investigated here, rotational and translational DH are not spatially decoupled, $1\over \tau_n$ should stay coupled to $\tau_\alpha$.

We next set out to determine if there were any morphological changes in the DH and to explore its connection to the breakdown of the SED relation. To identify DH, we picked the top $10\%$ orientationally most-mobile particles over $t^*$ and clustered them using a protocol followed earlier \cite{our_prl, yhan}. These clusters for $\phi = 0.76$ and $\phi =  0.79$ are shown in Fig.\ref{Figure1}d and e, respectively. We quantified the shapes of these clusters by finding the most probable number of orientationally fast nearest-neighbours for an orientationally fast particle, $P(NN^{\theta})$ \cite{RFOT_cluster_shape_2, yodh_prl}. A narrow $P(NN^\theta)$ that is peaked at $2$ implies string-like DH, while a broader distribution with a maximum beyond 2 signals the presence of more compact DH. Since small clusters will bias ${P(NN}^{\theta})$, we only consider cluster sizes $N^{\theta} \geq 4$ to quantify their morphology. Consistent with the predictions of the RFOT theory \cite{RFOT_cluster_shape_1, RFOT_cluster_shape_2}, we find that DH for the orientational DOF become more compact with supercooling (Fig. \ref{Figure1} d-f). Remarkably, the morphological change in DH from string-like to compact also coincides with the $\phi$ beyond which $D^\theta$ shows signatures of a breakdown of the SED relation (Fig. \ref{Figure1}c and f).  

\begin{figure}[htbp]
\includegraphics[width=0.6\textwidth]{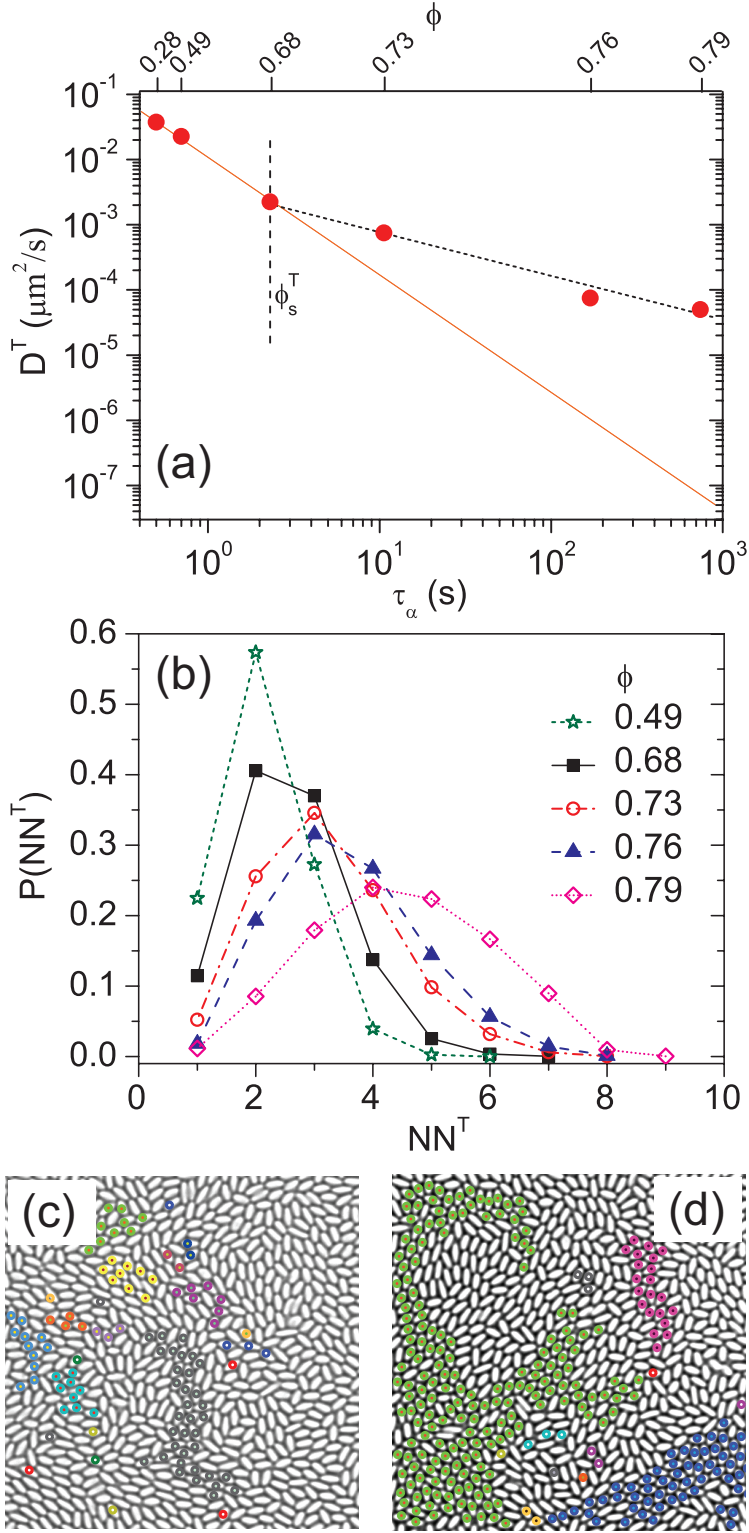}
\caption{(Color online). (a) Translational diffusivity $D^T$ versus the structural relaxation time $\tau_{\alpha}$. The lines in (a) show $\tau_\alpha^{-1.8}$ (solid) and $\tau_\alpha^{-0.7}$  (dotted line) dependencies. The vertical dashed line represents the dynamic crossover area fraction $\phi_s^T = 0.68$. (b) Most probable number of translationally fast nearest-neighbours for a translationally fast particle $P(NN^{T})$ for different $\phi$s. Clusters of top $10\%$ translationally most-mobile particles for $\phi = 0.76$ (c), $\phi = 0.79$ (d). In (c) and (d), the colors correspond to distinct clusters.}
\label{Figure2}
\end{figure}

\begin{figure}[htbp]
\includegraphics[width=1\textwidth]{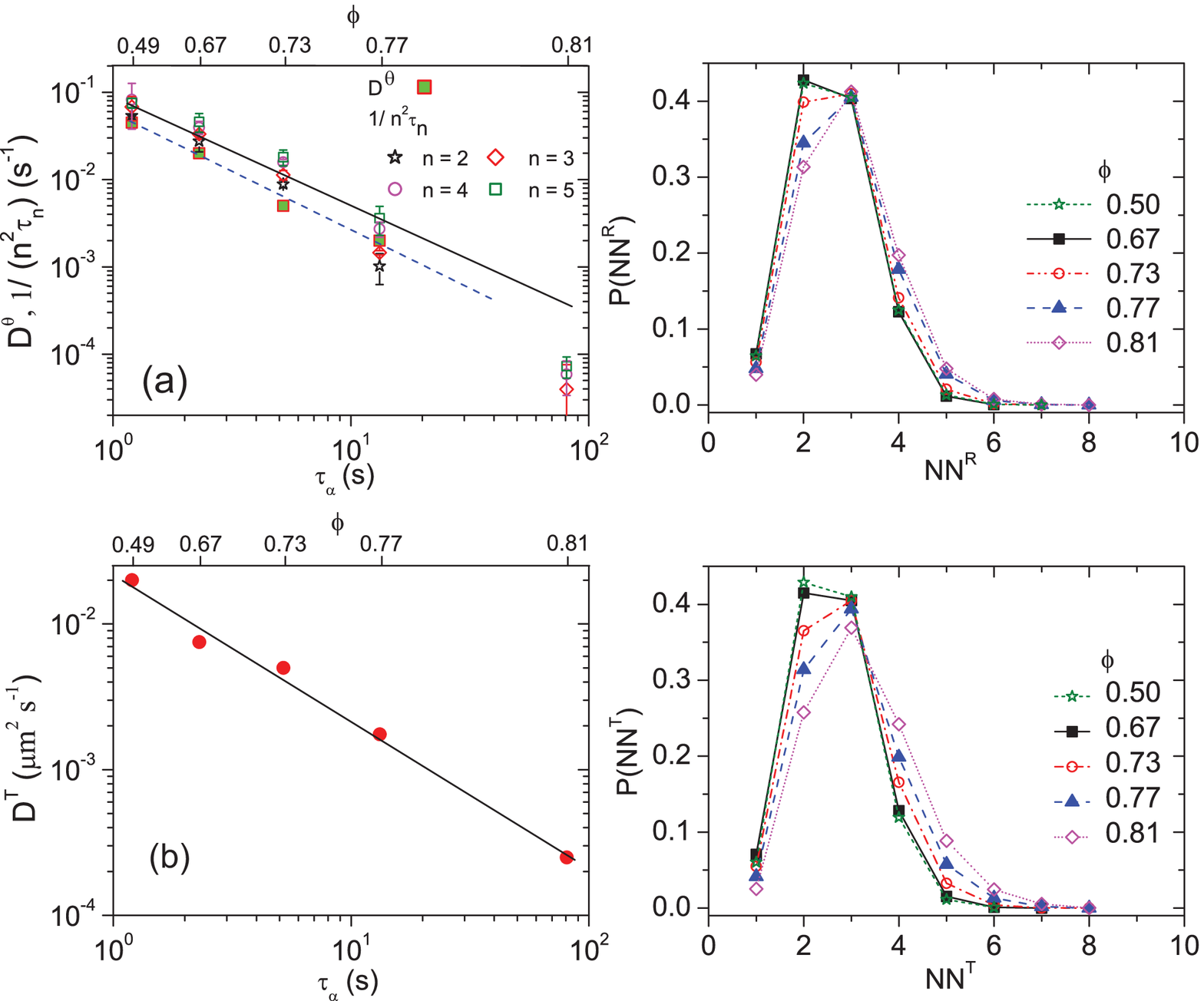}
\caption{(Color online). (a) Orientational diffusion coefficient $D^{\theta}$ and inverse of the $n^{th}$-order orientational relaxation time, $1/ {n^2 \tau_n}$ versus the structural relaxation time $\tau_{\alpha}$. The lines in (a) show $\tau_\alpha^{-1.3}$ dependencies. (b) Most probable number of orientationally fast nearest-neighbours for a orientationally fast particle $P(NN^{\theta})$ for different $\phi$s. (c) Translational diffusivity $D^T$ versus $\tau_{\alpha}$. The solid line in (c) shows ${\tau_{\alpha}}^{-1}$ dependency. (d) Most probable number of translationally fast nearest-neighbours for a translationally fast particle $P(NN^{T})$ for different $\phi$s.}  
\label{Figure3}
\end{figure}

Motivated by the above observations, the obvious next step was to examine if changes in the shape of DH in the translational DOF correlate well with the breakdown of the SE relation. Albeit the SE relation has been a subject of a large number of investigations \cite{DH_4, DH_5, breakdown_papers}, it is only recently that simulations have associated the breakdown of the SE relation with the morphological changes of DH \cite{szamel}. Figure \ref{Figure3}a shows the variation of $D^T$ with $\tau_\alpha$ for ellipsoids with purely repulsive interactions. Analogous to $D^{\theta}$, $D^T$ was extracted from long-time slope of the mean squared displacements, $\langle \Delta r^2 (t) \rangle = 4D^T t$ (see the Supplemental Material \cite{supplement}). For $\phi \leq 0.68$, $D^T \propto {\tau_ \alpha}^{-\xi}$ and again $\xi > 1$ (Fig. \ref{Figure2}a). For $\phi > 0.68$, however, the SE relation breaks down, and analogous to simulations \cite{eta_tau_alpha_1, stanley, breakdown_quantifiers_2}, we observed a fractional SE relation $D^T \propto {\tau_ \alpha}^{-\xi}$ with $\xi \approx 0.7$ (Fig. \ref{Figure2}a). Following our earlier line of analysis, we identified the most-probable number of translationally fast nearest-neighbours $P(NN^T)$ for a translationally fast particle. Strikingly, across $\phi = 0.68$, the DH become more compact and $P(NN^T)$ becomes progressively broader with $\phi$ (Fig. \ref{Figure2}b, c and d). This allows us to identify $\phi = 0.68$ with the dynamical crossover area fraction $\phi_s^T$.  These observations are not only in agreement with recent theoretical \cite{RFOT_cluster_shape_1} and numerical predictions \cite{szamel} but also bolsters our claim that the breakdown of SE and SED relation necessitates a change in the shape of the DH from string-like to compact.

To further our cause, it would suffice to show that in the \textit{absence} of changes in the shape of DH both SE and SED relations remain valid. In order to achieve this, we take recourse to findings from molecular dynamics simulation on attractive hard sphere glasses where a reentrant behaviour in the validity of the SE relation was observed \cite{zaccerelli}. While the SE relation breaks down for the repulsive as well as the strong attraction case, the dynamics was found to be faster at intermediate attraction strengths with the SE relation remaining valid even for the deep supercooling \cite{zaccerelli}. This study, however, did not probe the connection between the validity of SE relation and the nature of DH. In the context of ellipsoids, the introduction of small depletant molecules results in an anisotropic attraction that favors the lateral alignment of ellipsoids as opposed to tip-to-tip ones \cite{our_prl}. Although MCT for hard ellipsoids predicts a single glass transition for both the rotational and translational DOF for $\alpha < 2.5$ \cite{MCT_prediction}, recent experiments have observed that depletion attraction enhances pseudonematic ordering at intermediate attraction strengths \cite{our_prl}. Consequently, the orientational glass transition was found to precede the translational one and reentrant glass dynamics was observed only in the translational DOF (see the Supplemental Material \cite{supplement}. The dynamics for this system was observed to be fastest for an intermediate attraction strength of ${\Delta U \over {K_BT}} = -1.16$, and allowed access to $D^T$, $D^{\theta}$ and $\tau_n$ close to the glass transition. Moreover, the decay of $F_s(q,t)$ was found to be logarithmic close to $\phi_g^T$ and is indicative of its vicinity to the $A_3$ singularity (see the Supplemental Material \cite{supplement}). $\beta$ was once again found to decrease on approaching $\phi_g^T$. In contrast to the repulsive case, the growth of pseudonematic domains with $\phi$ hinders the relaxation of lower order orientational correlators to a greater degree than the higher order ones. Thus, while $1 \over {n^2\tau_n}$ collapses at low $\phi$ for all $n$, they show marked deviations on approaching $\phi^\theta_g$ (Fig \ref{Figure3}a). Also, a recent study on the same system has observed that orientational and translational DH are spatially decoupled at this interaction strength \cite{our_pnas} and we expect $1 \over \tau_n$ to progressively decouple from $\tau_\alpha$ as well. This is indeed the case here with the decoupling of lower order orientational correlators being more pronounced (Fig \ref{Figure3}a). The SED relation by the Einstein method remains valid even close $\phi^\theta_g$ with $D^{\theta } \propto {\tau_ \alpha}^{-\xi^E}$ with $\xi^E > 1$ (dashed line in figure \ref{Figure3}a). Most remarkably, consistent with the observed validity of the SED relation, $P(NN^{\theta})$ does not evolve with $\phi$ either (Fig. \ref{Figure3}b).  Further, we found that the SE relation remains valid even for deep supercooling with $D^T \propto {\tau_ \alpha}^{-\xi}$ with $\xi = 1$ (Fig. \ref{Figure3}c). Lending further strength to our findings $P(NN^T)$ does not show any appreciable change with increasing $\phi$ (Fig. \ref{Figure3}d). 

In conclusion, our study has helped establish a causal link between the  morphological evolution of DH and breakdown of the SE and SED relations. Our findings show unambiguously that, albeit DH are present at $\phi < \phi^{\theta,T}_s \sim 0.68$ in both the translational and orientational DOF, it is their change in the shape from string-like to compact that coincides with the breakdown of SE and SED relations. These results are in agreement with recent simulations \cite{szamel}.  Consistent with predictions of RFOT theory \cite{RFOT_cluster_shape_1}, DH become more compact on approaching the glass transition. While for ellipsoids with purely repulsive interactions, $D^T$ and $D^{\theta}$ decouple with $\tau_ \alpha$ beyond $\phi_s$, $1\over {\tau_n}$ does not. It has recently been proposed that $ \langle \Delta \theta^2(t) \rangle$, and consequently $D^\theta$, are overestimated due to the liberational motion of particles within cages \cite{eta_tau_alpha_1}. While this will result in an enhanced breakdown of the SED relation, the striking correlation observed between the decoupling of $D^\theta$ with $\tau_\alpha$ and the morphological changes of DH leads us to conclude that $D^\theta$ remains a reasonable measure of orientational dynamics. Turning on short-ranged attractive interactions, melted the glass and allowed us to access the translational and orientational dynamics even for deep supercooling. Although earlier studies, on the same system, have provided clear evidence for the presence of DH \cite{our_prl}, these heterogeneities continue to remain string-like and consequently both $D^T$ and $D^\theta$ stay coupled to $\tau_\alpha$. These observations further provide the first experimental confirmation of the validity of the SE relation along the $A_3$ singularity, a scenario that has remained untested even in supercooled liquids of attractive hard spheres. 

CKM thanks CPMU, JNCASR and RG thanks ICMS, JNCASR for financial support.

\begin{thebibliography}{10}
\providecommand*{\bibinfo}[2]{#2} \providecommand*{\eprint}[1]{#1}
\providecommand*{\url}[1]{#1}

\bibitem {DH_1}
G. Adam and J. H. Gibbs, J. Chem. Phys. \textbf{43}, 139 (1965).

\bibitem{DH_2}
M. D. Ediger, Annu. Rev. Phys. Chem. \textbf{51}, 99 (2000).

\bibitem{DH_3}
W. K.Kegel and A. van Blaaderen, Science \textbf{287}, 290 (2000); E. R. Weeks, J. C. Crocker, A.C. Levitt, A. Schofield, and D. A. Weitz, Science \textbf{287}, 627 (2000).

\bibitem{DH_4}
L. Berthier, G. Biroli, J.-P. Bouchaud, L. Cipelletti, and W. van Saarloos, \textit{Dynamical heterogeneities in glasses, colloids and granular materials}, Oxford University Press, Oxford, 2011.

\bibitem{DH_5}
J. A. Hodgdon and F. H. Stillinger, Phys. Rev. E \textbf{48}, 207 (1993); F. H. Stillinger, J. A. Hodgdon, Phys. Rev. E \textbf{50}, 2064 (1994).

\bibitem{einstein}
A. Einstein, \textit{Investigations on the Theory of Brownian Motion} (Dover, New York, 1956).

\bibitem{debye}
P. Debye,\textit{ Polar Molecules} (Dover, New York, 1929).

\bibitem{breakdown_papers}
E. R{\"o}ssler, Phys. Rev. Lett. \textbf{65}, 1595 (1990); L. Xu, F. Mallamace, Z. Yan, F. W. Starr, S. V. Buldyrev, and H. E. Stanley, Nature Physics \textbf{5}, 565 (2009); Gilles Tarjus and Daniel Kivelson, J. Chem. Phys \textbf{10}3, 3071 (1995); P. Kumar, S. V. Buldyrev, S. R. Becker, P. H. Poole, F. W. Starr, and H. E. Stanley, Proc. Natl. Acad. Sci. U.S.A. \textbf{104}, 9575 (2007).

\bibitem{breakdown_quantifiers_1}
M. T. Cicerone and M. D. Ediger, J. Chem. Phys. \textbf{103}, 5684 (1995). 

\bibitem{breakdown_quantifiers_3}
X. Xia and P. G. Wolynes, Phys. Rev. Lett. \textbf{86}, 5526 (2001).

\bibitem{breakdown_quantifiers_2}
S. Sengupta, S. Karmakar, C. Dasgupta, and S. Sastry, J. Chem. Phys. \textbf{138}, 12A548 (2013).

\bibitem{dimension_dependence_1}
J. D. Eaves and D. R. Reichman, Proc. Natl. Acad. Sci. U.S.A. \textbf{106}, 15171 (2009).

\bibitem{szamel}
E. Flenner, H. Staley, and G. Szamel, Phys. Rev. Lett. \textbf{112}, 097801 (2014); E. Flenner and G. Szamel, J. Chem. Phys. \textbf{138}, 12A523 (2013).

\bibitem{RFOT_cluster_shape_1}
J. D. Stevenson, J. Schmalian, and P. Wolynes, Nature Physics \textbf{2}, 268 (2006).

\bibitem{RFOT_cluster_shape_2}
K. H. Nagamanasa, S. Gokhale, A. K. Sood, and R. Ganapathy, arXiv preprint arXiv:1408.5485 (2014).

\bibitem{yodh_prl}
Z. Zhang, P. J. Yunker, P. Habdas, and A. G. Yodh, Phys. Rev. Lett. \textbf{10}7, 208303 (2011).

\bibitem{stanley}
S. R. Becker, P. H. Poole, and F. W. Starr, Phys. Rev Lett \textbf{97}, 055901 (2006); M. G. Mazza, N. Giovambattista, H. E. Stanley, and F. W. Starr, Phys Rev. E \textbf{76}, 031203 (2007).

\bibitem{SED_valid_Debye}
I. Chang, F. Fujara, B. Geil, G. Heuberger, T. Mangel, H. Sillescu, J. Non-Cryst. Solids \textbf{172-174}, 248 (1994); M. T. Cicerone, F. R. Blackburn, and M. D. Ediger, J. Chem. Phys. \textbf{102}, 471 (1995). 

\bibitem{eta_tau_alpha_1}
S. H. Chong, W. Kob, Phys. Rev. Lett. \textbf{102}, 025702 (2009). 

\bibitem{weeks}
K. V. Edmond, M. T. Elsesser, G. L. Hunter, D. J. Pine, and E. R. Weeks, Proc. Natl. Acad. Sci. U.S.A. \textbf{109}, 17891 (2012).

\bibitem{tracers}
R. Zangi, S. A. Mackowiak, and L. J. Kaufman, J. Chem. Phys. \textbf{126}, 104501 (2007); S. A. Mackowiak, J. M. Noble, and L. J. Kaufman, J. Chem. Phys. \textbf{135}, 214503 (2011); D. B. Hall, D. D. Deppe, K. E. Hamilton, A. Dhinojwala, and J. M. Torkelson, J. Non-Cryst Solids \textbf{235-237}, 48 (1998).

\bibitem{our_prl}
C. K. Mishra, A. Rangarajan and R. Ganapathy, Phys. Rev. Lett. \textbf{110}, 188301 (2013).

\bibitem{supplement}
See the Supplemental Material for the experimental details, phase diagram of the colloidal ellipsoids of $\alpha = 2.1$, signature of $A_3$ singularity and extracting of $D^T$ and $D^{\theta}$.

\bibitem{yhan}
Z. Zheng, F. Wang, Z. Zheng, and Y. Han, Phys. Rev Lett. \textbf{107}, 065702 (2011).

\bibitem{bagchi}
P. P. Jose, D. Chakrabarti, and B. Bagchi, Phys. Rev. E \textbf{73}, 031705 (2006).

\bibitem{eta_tau_alpha_2}
R. Yamamoto and A. Onuki, Phys. Rev. Lett. \textbf{81}, 4915 (1998); F. Mezei, W. Knaak, and B. Farago, Phys. Rev. Lett. \textbf{58}, 571 (1987).


\bibitem{horrowell}
D. N. Perera and P. Harrowell, Phys. Rev. Lett. \textbf{81}, 120 (1998).

\bibitem{zaccerelli}
A. M. Puertas, C. D. Michele, F. Sciortino, P. Tartaglia, and E. Zaccarelli, J. Chem. Phys. \textbf{127}, 144906 (2007).

\bibitem{MCT_prediction}
M. Letz, R. Schilling, and A. Latz, Phys. Rev. E \textbf{62}, 5173 (2000).

\bibitem{our_pnas}
C. K. Mishra, K. H. Nagamanasa, R. Ganapathy, A. K. Sood, and S. Gokhale, \textit{in press} Proc. Natl. Acad. Sci. U.S.A.; arXiv preprint arXiv:1408.0343, (2014).

\end{thebibliography}
\end{document}